# cniCloud: Querying the Cellular Network Information at Scale

Wenguang Huang[1], Chang Zhou[1], Yuanjie Li[2]

[1]Shanghai Jiao Tong University
Department of Electric Engineering
Shanghai, China
[2]University of California, Los Angeles
Computer Science Department
Los Angeles, California, United States
{blueskygundam,lemonbirdy}@sjtu.edu.cn,yuanjie.li@cs.ucla.edu

## ABSTRACT

This paper presents cniCloud, a cloud-based platform for mobile devices to share and query the fine-grained cellular information at scale. cniCloud extends the single-device cellular analytics via crowdsourcing: It collects the fine-grained cellular network data from massive mobile devices, aggregates them in a cloud-based database, and provides interfaces for end users to run SQL-like query over the cellular data. It offers efficient and responsive processing by optimizing the database storage, and adopting the domain-specific optimizations. Our preliminary deployments and experiments validate its feasibility in performing crowdsourced analytics.

## 1 INTRODUCTION

Cellular networks (3G, 4G LTE and upcoming 5G) have been an integral part of our daily life. As large-scale wireless infrastructure, they provide "anywhere, anytime" Internet access to billions of mobile users, contributing to 69% of mobile data traffic in 2016 [1]. Understanding and analyzing the cellular network behaviors is thus critical for not only network operators, but also researchers and developers.

An obstacle for researchers and developers to understand the cellular network is its "black-box" nature *at scale*. Besides basic status and functions, the mobile devices (e.g. smartphones and tablets) only have limited access to the coarse-grained cellular network information in the usual scenarios. Recent efforts (notably MobileInsight [2]) seek to address this by enabling the in-device open access to the fine-grained cellular information. However, such approach is based on single device, thus unable to comprehensively analyze the large-scale infrastructure with diverse, dynamic and heterogenous behaviors among distributed network nodes and operators.

In this work, we seek to address this by exploring the crowdsourcing approach. While each device has limited cellular information, the aggregation of cellular information from massive devices would offer more valuable cellular network information. A platform for this purpose would help researchers and developers better understand the large-scale, complex cellular network behaviors.

In achieving this, we face two main challenges. First, such solution should be scalable to the huge cellular data volume from numerous mobile devices. Second, the aggregation and analytics of the massive cellular data should be efficient in terms of processing speed and system overhead.

We present cniCloud, a cloud-based platform for the collection and query of fine-grained cellular information at scale. cniCloud collects the fine-grained cellular network data from massive mobile devices, aggregates them in a cloud-based database, and provides interfaces for end users to run SQL-like query over the cellular data. The whole query and database system is based on Spark [3], which supports scale-out property. It offers efficient and responsive processing by optimizing the database storage, and adopting the domain-specific optimizations. Our preliminary deployments and experiments validate its feasibility in performing queries covering multiple phones at large areas, while retaining acceptable overheads.

A preliminary version of cniCloud is available at [4].

## 2 CELLULAR NETWORK PRIMER

The cellular networks are currently the largest wireless infrastructure that offers "anywhere, anytime" network services to mobile users. Figure 1 (right) illustrates its general architecture. It consists of a radio access network (RAN) and a core network. The RAN is provisioned by the base stations, and provides wireless access to the mobile clients (e.g. smartphones). The core network connects the RAN to the Internet. To offer wide-area network access, both the radio access and core network span on large geographical areas.

Similar to the Internet, the cellular network adopts a set of protocols to offer critical network functions (including wireless, mobility and data session management). Figure 1 (left) illustrates the protocol stack. To enable the wireless communication between the client and the radio base station, cellular network defines physical and link-layer protocols, including PHY, MAC, RLC (Radio Link Control) and PDCP (Packet Data Convergence Protocol). On top of it, a set of control plane signaling protocols are defined, including (1) the radio resource control (RRC) protocol for radio resource allocation and connection management; (2) the mobility management (MM) protocol for client location update and mobility support; and (3) the session management (SM) protocol for voice/data session establishment and maintenance. The messages exchanged within these protocols carry rich information about the cellular network, such as the protocol status dynamics, configurations, and operation policies.

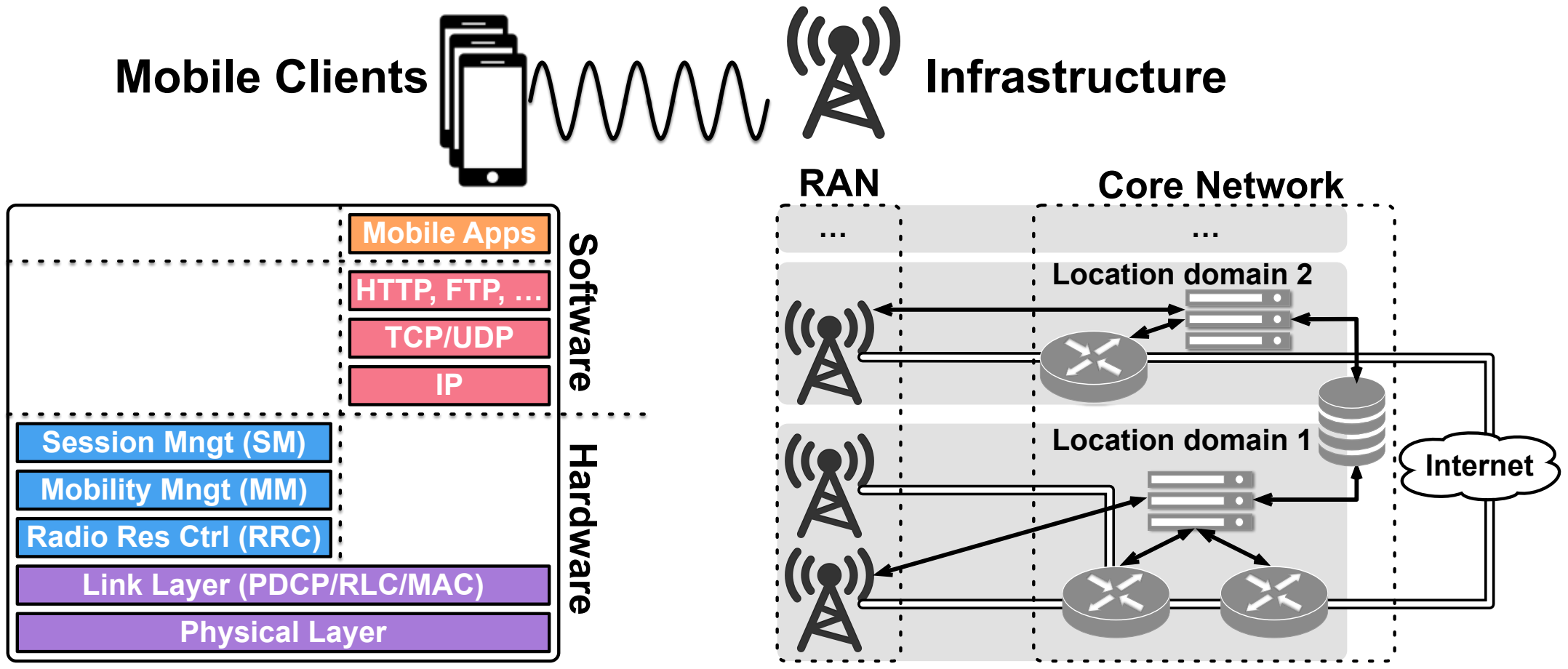

**Figure 1: 4G LTE cellular network architecture, and protocol stacks.**

# 3 MOTIVATING SCENARIOS

To understand the large-scale network behaviors, it is beneficial to enable the cellular data sharing among mobile devices. We next consider three scenarios empowered by this.

**Characterizing the cellular message patterns:** Consider how to characterize the patterns of cellular message. In particular, we may want to learn the distribution of the distinct message counts. Such distribution offers basic information for the cellular network operations. To do it, one way is to collect, classify and count the cellular messages inside a single test phone. However, such results may be biased with the noises of each phone's specific usage scenarios. Instead, by aggregating the results from multiple mobile devices, such bias could be mitigated.

**Comparing the cellular network operators:** For better performance and network accessibility, the mobile users may want to compare the cellular network operators. One example is Google Project Fi [5], which allows the smartphones to use the multiple carrier networks (e.g. , T-Mobile and Sprint). By ranking the available operators, the Fi-empowered smartphones could select the best cellular networks and improve user experience. By sharing the cellular-level data, mobile devices can collaboratively improve the ranking accuracy and thus their performance.

**Distributions of specific cellular parameters:** Some cellular parameters are critical for device-perceived network behaviors. For example, the radio resource control (RRC) protocol supports the power saving via periodical sleeping mode. The sleeping interval is determined by the configurable timers (e.g. $T_{shortDRX}$ specified by [6]). These timers are configured by the serving base station, and vary among network nodes. The distribution of such timer value would thus require cellular data from multiple devices at different geographical areas.

# 4 CNICLOUD DESIGN

This section presents the design of cniCloud. cniCloud's goal is to enable structured query (e.g. SQL-like) of fine-grained cellular network information at scale. This requires three functions: (1) *Log collection:* cniCloud should collect fine-grained cellular network data from massive mobile devices; (2) *Data management:* Given the rich cellular data, cniCloud should properly manage them to facilitate the query efficiently; (3) *Query:* cniCloud should provide easy-to-use interfaces for structured query. In achieving them, cniCloud should still retain scalability and high efficiency.

Figure 2 illustrates the overview of cniCloud. It includes mobile devices that share the cellular data, and a centralized database system inside the cloud. On the mobile device side, cniCloud extends the sing-device fine-grained cellular data collection with sharing functions (§4.1). Inside the cloud, cniCloud constructs the database to store and aggregate cellular data (§4.2), and adopts SQL-like interfaces for the query (§4.3). It also provides a front-end website server for user queries. We next elaborate each.

## 4.1 Log Collection From Massive Phones

The first step for cniCloud is to collect the cellular network data from massive mobile devices. Unfortunately, the state-of-art OS APIs do not suffice for this purpose: It only offers basic status and functions of cellular networks, which cannot support the fine-grained cellular information query.

**MobileInsight primer:** MobileInsight [2] is a user-space mobile app that collects and analyzes the fine-grained cellular network information on the off-the-shelf smartphones. To collect the low-level cellular information, MobileInsight does not rely on the legacy mobile OS APIs. Instead, it explores an alternative side channel (*diagnostic mode*) across the hardware chipset and the software, and exposes raw protocol control/data messages to the user space. This feature meets cniCloud's basic demands for cellular information gathering.

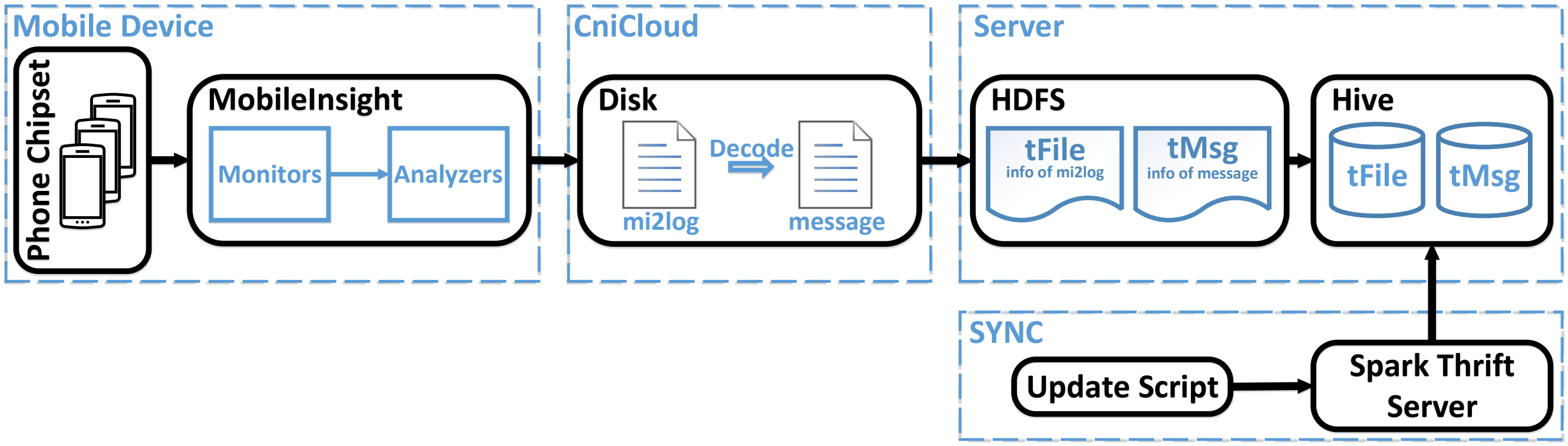

**Figure 2: The system overview of cniCloud.**

**Extensions for data sharing:** To enable the sharing of fine-grained cellular data, cniCloud extends MobileInsight as follows. On the phone side, it develops a MobileInsight plugin to collect the cellular data, and upload them to the cniCloud cloud. On the cloud side, cniCloud adopts a set of scripts for the log gathering, classification and metadata construction. To facilitate the data classifications on the cloud, all the updated logs are renamed as follows:

```
cniCloud_TIMESTAMP_LOC_MODEL_OPERATOR.log
```

where `TIMESTAMP` and `LOC` are the time and GPS locations the log was collected, respectively. `MODEL` is the phone model, and `OPERATOR` is the cellular network operator the phone is using. In this way, cniCloud can support temporal/spatial cellular data analytics, and aggregate cellular logs by phone models and/or network operators.

## 4.2 Data Management

With rich dataset from the massive phones, the next step for cniCloud is to properly manage these data. We build cniCloud's database on with Spark SQL [3], an efficient and scalable in-memory data processing engine using SQL-like and key-store value interfaces.

**Metadata constructions:** To facilitate the queries, cniCloud manages the cellular data with two metadata tables: `tFile` and `tMsg`. They are declared as follows:

```
CREATE TABLE tFile (
    Filepath VARCHAR[], Phone VARCHAR[],
    Carrier VARCHAR[], Time TIMESTAMP);

CREATE TABLE tMsg (
    Filepath VARCHAR[], Time TIMESTAMP,
    MsgType VARCHAR[], MsgHash VARCHAR[],
    MsgPath VARCHAR[], LineNo VARCHAR[]);
```

The `tFile` table stores the information of the original MobileInsight logs. It has attributes including file path, phone, carrier and timestamp. Table `tMsg` stores the information of the decoded message log files. It has attributes including file path, timestamp, message type, message hash, message path and row numbers. The message log files stores the detailed information how the underlying protocol operates. They are made up of multiple-nesting of key-value pairs which is a complex structure. In this way, our design of database management system should not only focus on support of relational database, but also support of key-value database.

**Optimizations for query efficiency:** cniCloud seeks to support fast queries of the cellular data at scale. The key is to optimize how the data is stored and organized. To this end, cniCloud adopts two optimizations: *index-based metadata*, and *in-memory processing*.

∘ *Index-based metadata:* To support the structured query, cniCloud should store the cellular data in a structured way. One approach is to directly store the data into the database. This turns out to be slow: Our experiments find that, cniCloud's query speeds degrade significantly with the growing database size. Instead, we extend `tMsg` with two extra log indexes: `MsgPath` and `LineNo`. For each cellular message, `MsgPath` specifies the original log path, and `LineNo` specifies the index of the message in this file. In this way, cniCloud does not need to store the entire cellular data into the database. The cost is that, every query needs to process the raw cellular data. Such delay turns out to be tolerable, compared with the latency caused by database scaling.

∘ *In-memory processing:* In building cniCloud, a major performance bottleneck is the storage I/O. Our early version of cniCloud used MySQL, which reads data directly from disk and process the query operations through disk I/O. This turns out to be slow due to the disk I/O inefficiency (see §5).

To this end, our current cniCloud uses SparkSQL [3], an efficient and scalable in-memory data processing engine using SQL-like and key-store value interfaces. SparkSQL supports both relational database and key-value database well and it deals with query operations through memory which effectively omit the problem disk IO brings. SparkSQL also optimizes the query on its distributed realization. After a detailed comparison between SparkSQL and query function of other database systems, the paper nd that SparkSQL has its unique advantages. Therefore, the paper adopts SparkSQL as its database query component.

The query operations of SparkSQL are based on a data structure called `Dataframe`. `Dataframe` has a new concept called Schema. It is like the table structure in MySQL. The schema records the field names that each column belongs to. To create a `Dataframe`, we should firstly define the Schema, then fill it with the table content. cniCloud further uses Hive [7] to cache the table which is also capable of putting the tables into memory. It is mainly determined by the characteristic of Spark Thrift Server. And the combination of these components is accomplished by Spark Thrift Server.

**Scale-out support:** To tackle the growing cellular data, cniCloud's database support scale-out property. This is empowered by Spark, which support HDFS-like distributed file system to store the raw cellular data. By adding more storage servers, cniCloud can hold more cellular data and retain the same query efficiency. Note that, the metadata files are stored inside memory rather than on the disk for efficiency.

## 4.3 Querying Interfaces

With cniCloud's database, users can query the cellular information collected from massive phones in large areas. We have built a front-end website [4] to support this query. In the following, we first exemplify the queries by revisiting the scenarios in §3, and present the issues and solutions in building the query interface.

**Examples:** We consider the scenarios in §3.

*1. Characterizing the cellular message patterns:* With `tMsg` metadata, such query becomes straightforward:

```
SELECT MsgType, count(*) FROM tMsg
GROUP BY MsgType;
```

In this example, we count the cellular messages, and group them by the message types. It readily offers us the distribution of the cellular message patterns. Note that such query aggregates data from all the phone models under various scenarios, thus mitigating the biases due to specific usages.

*2. Comparing the cellular operators:* cniCloud supports the query based on cellular operators. This allows users to characterize each operator, and compare different operators. Consider T-Mobile as one example. To query its characteristics of radio resource control, an intuitive query would be:

```
SELECT count(*) FROM tMsg
ON tMsg. Filepath = tFile.Filepath
WHERE tFile.Carrier = "T-Mobile"
AND tMsg.MsgType = "4G_LTE_RRC";
```

*3. Distributions of specific cellular parameters:* In cniCloud, querying a cellular parameter currently takes two steps. First, the users query the cellular messages that carry this parameter. In the example of $T_{shortDRX}$ in §3, users should query 4G RRC messages since they carry this parameter:

```
SELECT * FROM tMsg
WHERE tMsg.MsgType = "4G_LTE_RRC";
```

Such query returns the raw message contents. Then the users can further query $T_{shortDRX}$ out of these raw data. We are currently developing an alternative and more efficient query approach with the key-value store, as we will discuss in §6.

**Building the web-based query interface:** While cniCloud uses SparkSQL, SparkSQL does not have a complete interface, like MySQL, for web server to visit. To this end, it uses the framework based on Spark Thrift Server. Spark Thrfit Server provides us with a java jdbc interface which makes the database could be visited in ways like MySQL. It can also open a process running SparkSQL and listen to it for interaction. Using Spark Thrift Server can meet both of the requests. So the paper develops a framework as following:

◦ *Fault tolerance:* Website should be resilient to exceptions and user-made mistakes, which is also called robustness. As a database query website, our task is to check the input users input to see whether they are legal. The paper divide the error detection into two parts. The first part is designed to deal with blank instructions like space, tab or just empty. The second part is designed to deal with illegal instructions. The paper utilizes the error detection mechanism in SparkSQL. If some instruction is wrong, SparkSQL executes it and would call back an exception. The Spark Thrift Server can listen to this exception and the server can get error message through JDBC interface. In this way, users can get the same exact error message for exception as they directly using SparkSQL.

◦ *File Download:* Given the huge data volume, cniCloud supports the users to download the queries as a summary report. File download function offers users with approach to log files. After they get the query result and want to go deep into the content and structure of log files, they can download these files from the website.

The paper accomplishes another servlet class to deal with downloading. The idea is to read the file into input stream from the local and send it to output stream of the server. The detailed procedure is as following: (1) The servlet instance gets the filename to be downloaded and ensure its absolute path. (2) The instance defines FileInputStream and read files into it. (3) The instance checks the type of the file and fills the head of the request with it. (4) The instance obtains OutputStream from the response. (5) The instance sets the size of transferring buffer. Then it continually reads the content from FileInputStream and writes into OutputStream until the end of the file. (6) The instance closes FileInputStream and OutputStream.

# 5 PRELIMINARY RESULTS

In this section, we report our early results in cniCloud's efficiency in performing queries at scale, and its system overheads. We run experiments to test the performance and overhead of our systems and compare with other systems.

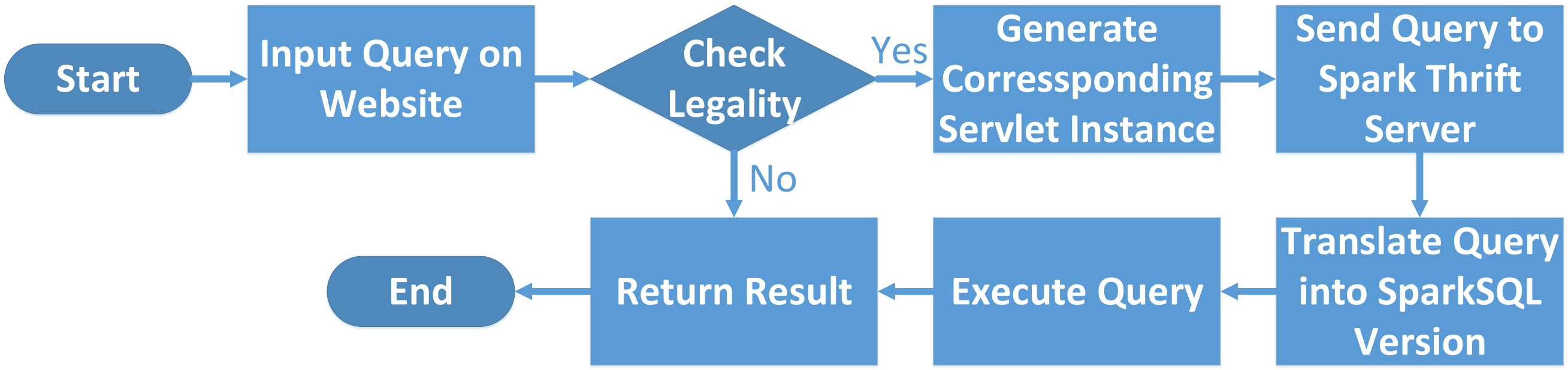

**Figure 3: Flowchart: User Querys Website**

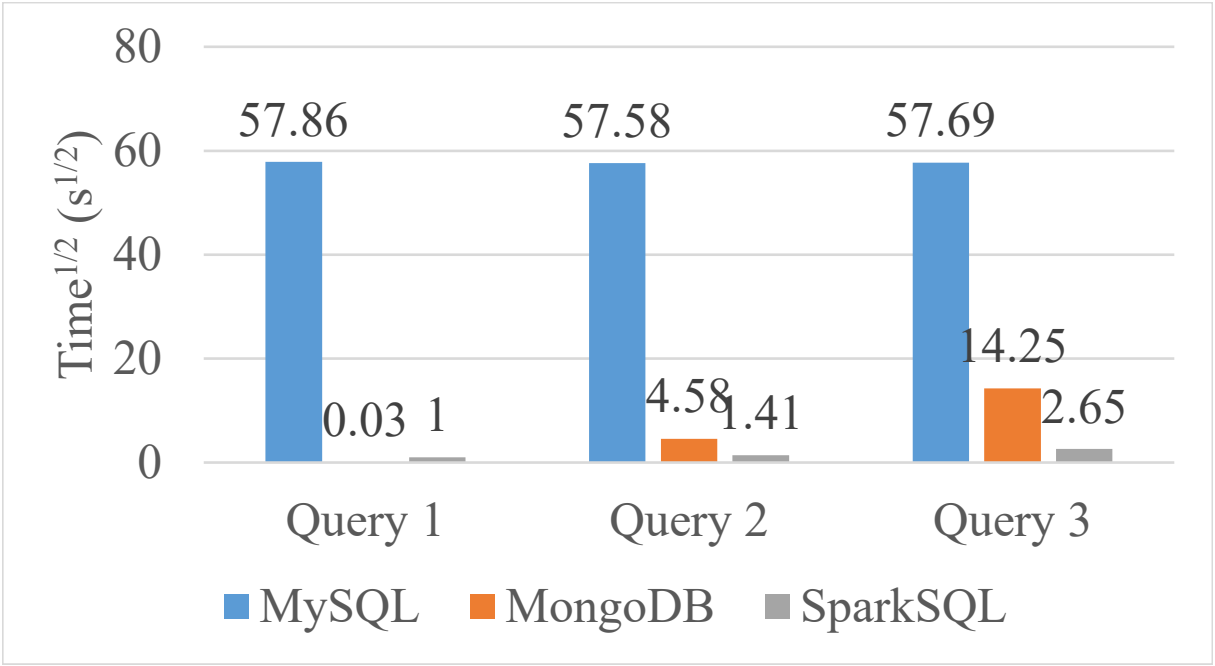

**Figure 4: Performance Test among Database System**

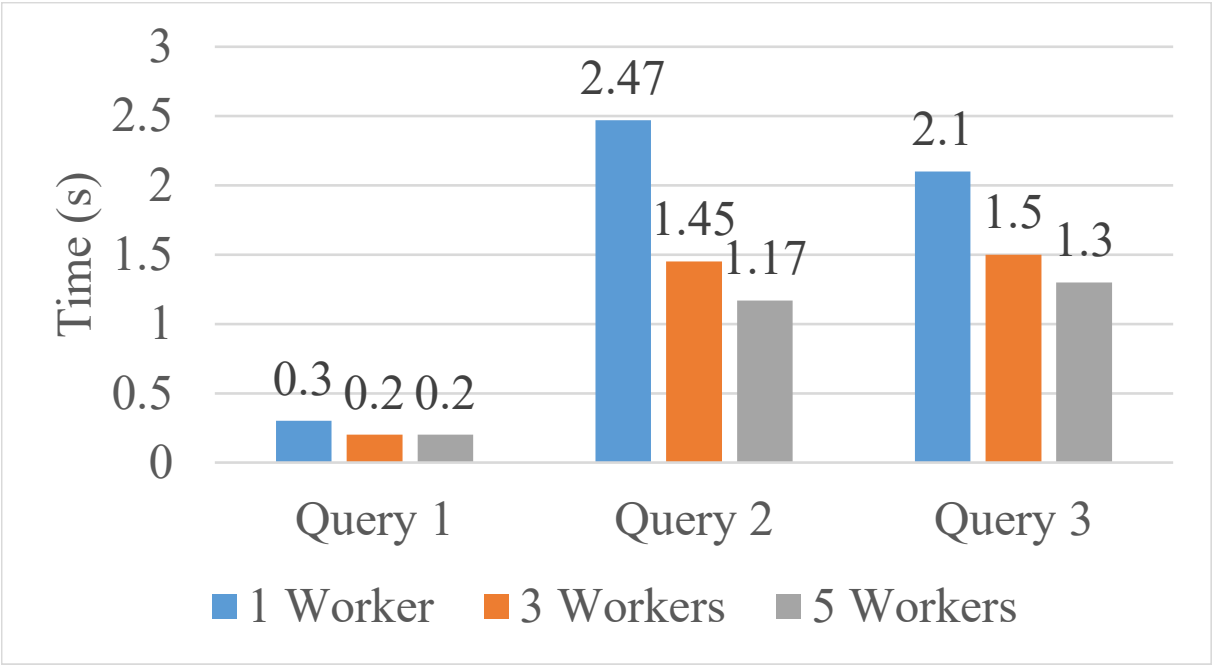

**Figure 5: SparkSQL Performance Test Using Different Workers**

**Dataset:** We run cniCloud over a MobileInsight dataset [8] from the 19-month period from 07/31/2015 to 02/28/2017. This dataset includes traces from 23 phones and seven Android models (Google Pixel, Huawei Nexus 6P, Motorola Nexus 6, Samsung Galaxy S4/S5, LG Optimus 2, and LG Tribute) from four U.S. carriers (AT&T, T-Mobile, Verizon, Sprint). It carries 1,222,673 4G/3G signaling messages in total, and results in 9.6 GB `tFile` and 20 GB `tMsg` table in total.

## 5.1 Query Performance

We conduct two tests. The first test compares the performance of query between cniCloud and other possible solutions. We tried three different database query system which are MySQL, MongoDB and SparkSQL respectively. In the experiment, we run three representative queries (shown in §4.3) to assess their execution time. We put the result into a histogram. In Figure.4, the y-coordinate represents the square root of execution time. We can see SparkSQL runs queries the most quickly which shows a great advantage over the other two database query systems. Such high speed of query satisfies the need of our design target.

Notice that when using Query 1, MongoDB runs 0.03s which defeats SparkSQL with a large gap. The reason is the query itself. The result is the total number of rows which MongoDB keeps this value in its system. MongoDB executes Query 1 without any real query operations and therefore it can give back the result in such short time.

The second test is to watch the query performance of SparkSQL using different number of workers. We choose 1, 3, 5 workers to run SparkSQL and run queries to observe the changes when the number of workers decrease. In Figure5, there is an great speed-up when we increase the number of workers from 1 to 3. This great improvement on speed gives the credit to the mechanism of distributed computation. When the number of workes increases, the computation can take more advantage that the distributed mechanism of SparkSQL brings. This adaptability contributes to the speed-up of queries.

The tendency is slowed when we keep increasing the number of workers. In our analysis, the main reason is the scale of the database. Since no more worker is required to do jobs, the speed gets to its bottleneck. In other hand, when the databases scale out, we can increase the number of workers in order to keep performance well. And that is the great potential our system could bring.

## 5.2 Resource Overhead

In this section, we measure the resource overhead of the system to observe the resource consumption when we enlarge the scale of our system. We keep inputing a group of instructions into the system in a short period. Then we change the number of workers and do the same test. We record the result and draw a graph represent the consumption of CPU and memory. In Figure.7 and Figure.6, we can see that CPU is always full-loaded when executing a query. When executing the same query operation, the more workers SparkSQL have, the shorter the duration lasts. And we can also find that the more workers SparkSQL have, the less memory and CPU each worker would use.

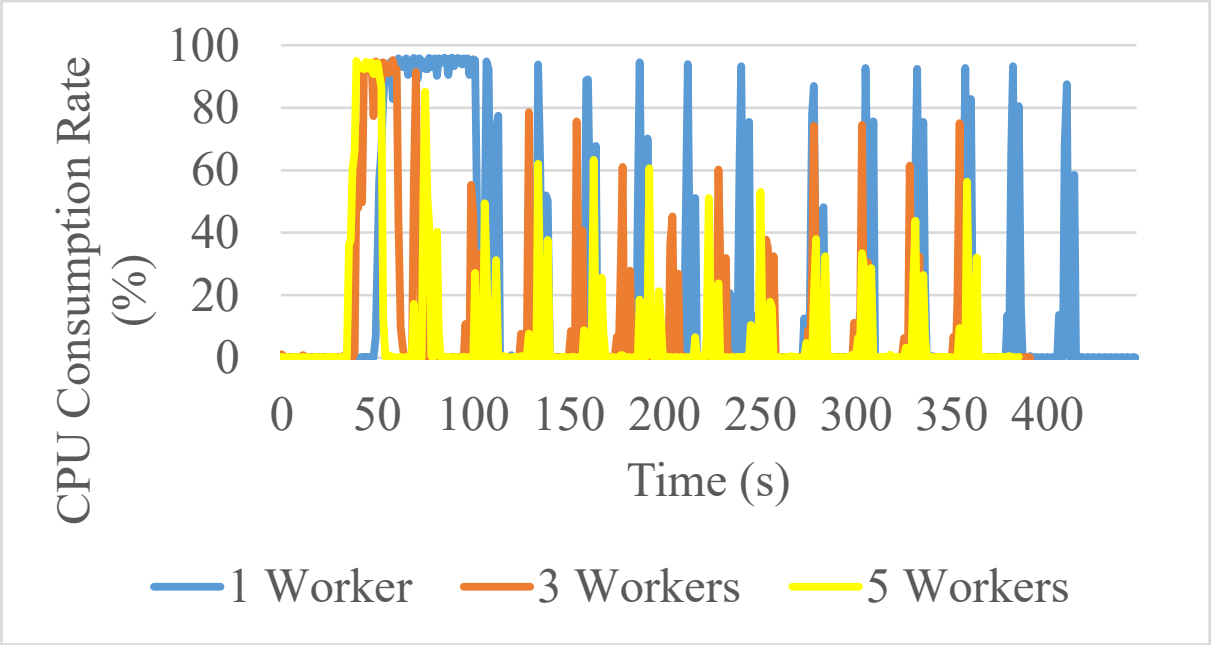

**Figure 6: CPU Overhead Test**

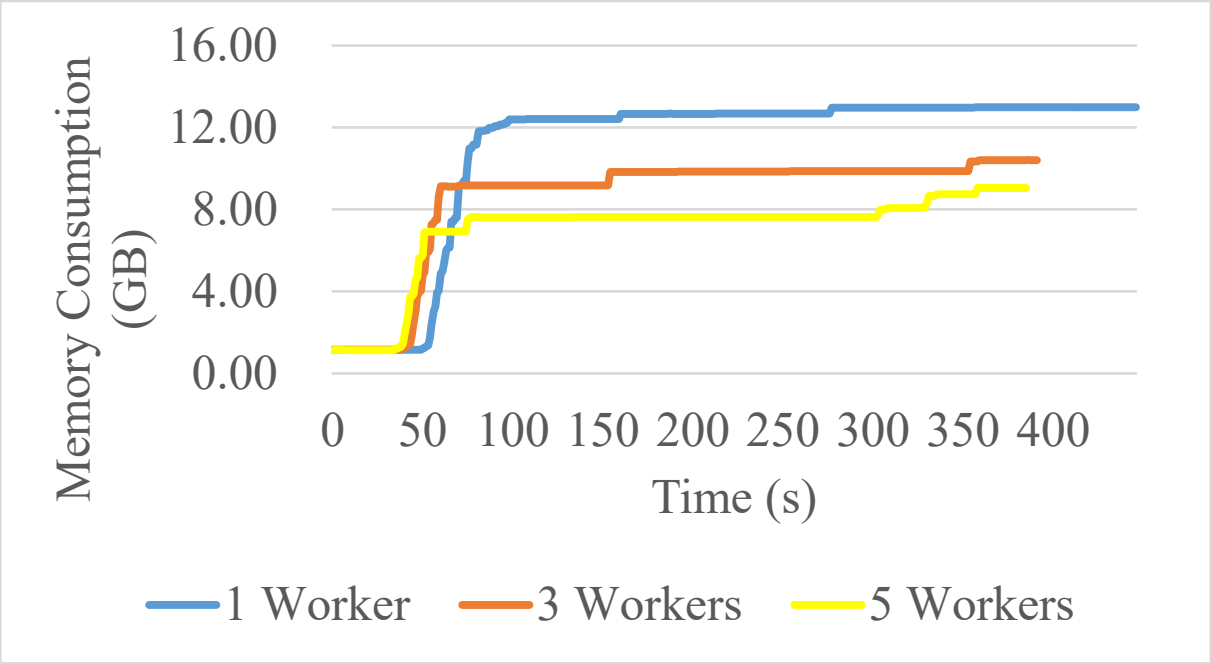

**Figure 7: Memory Overhead Test**

That also shows the scalability of cniCloud. When we want to enlarge the scale of the server, the consumption of each worker will not increase but decrease. In this way, we could put our system into use when dealing with a database with larger scale. On the other hand, such scablilty can bring a strong practicability. When we use more workers, the consumption of each worker decreases. Developers can make use of devices of low configuration to build the system with similar performance.

# 6 DISCUSSION

cniCloud is still at early stage. Our ongoing work aims to improve the organization of the original low-level cellular network data, and add more function for the query tool. Specifically, we are developing a new table named tDetail, which contain more message content of the original data. And based on tDetail table, we are developing a pattern search function for the query tool.

**Parameter query via key-value store:** The first ongoing work is the organization of tDetail table. Previously, we have talked about the design and realization of the tFile and tMsg table. With these 2 tables, reseachers can easily locate the log file and log message they need. However, reseachers still need to download the log files individually and read the messages by themselves. Different with tMsg table, tDetail table will contain more details of the message, including each component of a real message content. With tDetail, researchers do not have to download the log files, but they can read the message content directly on the cniCloud website.

There are two major challenges of realizing tDetail table: the resource requirement and the data structure. As you can see, building tDetail table means that we have to put the whole data into the memory, which is a burden to the limited memory resource. However, this challenge is easy to solve by Spark distributed memory management and adding enough memory into the cluster. The second problem, data structure problem, is more tricky. Then length of the celluar network message is not settled, which means that the length of different message could vary a lot from others. So we could not simply use a relational database table to store the message data. Instead, key-value database table is the optimized choice for this problem. So now we are using Spark API to try to store the tDetail in a key-value structure.

**Procedure pattern match:** The second ongoing work is the development of pattern search. We can see that there are many certain patterns of messages when some event happen. For example, when a network failure due to voice QoS misconfigurations happens, there will be a certain pattern of error messages generated in the log file. By searching the certain pattern in the whole data, we can locate more and more similar messages, which is useful for further cellular network researches. Futhermore, we plan to add user-defined pattern function in the future. So, we can not only search the messages with some given certain pattern, but we can also define our own message pattern to organize the message.

# 7 RELATED WORK

The past few years have witnessed a proliferation of experimental testbeds for cellular network research. These efforts span on three dimensions: (1) **SDR-based approaches:** This includes various software-defined radio (SDR) platforms (LTE-Eye [9], WARP [10], Sora [11], Tick [12], etc) equipped with software-defined 4G/3G cellular protocol stacks (OpenAirInterface [13], srsLTE [14], OpenLTE [15], openEPC [16], etc). cniCloud complements these efforts by leveraging the commodity phones' data for network research. (2) **Single-client approaches:** This includes various in-phone cellular network monitoring tools, such as MobileInsight [2], Mobilyzer [17], MobiPerf [18], to name a few. cniCloud moves one step further than these efforts, by crowdsourcing the cellular network information from mobile clients at scale. (3) **Multi-client approaches:** Some platforms crowdsource the coarse-grained cellular network information from various phones, such as OpenSignal [19] and PhoneLab [20]. Instead, cniCloud enables queries for fine-grained cellular network information at scale, by using the below-IP network data.

cniCloud is inspired by various generic distribute computing platforms, such as Spark [3]), Hadoop [21]. It differs from these efforts since it focuses on the cellular network analytics. Recent efforts [22, 23] seek graph-based approach to model and process the cellular network data from the infrastructure. cniCloud differs from them since it is based on structured query (SQL) and client-side cellular data.

## 8 CONCLUSION

In this paper, we report our first effort of crowdsourcing fine-grained cellular information at scale. we present cniCloud, a cloud-based platform to enable the query of cellular information from massive phones. We have described the architecture and working procedure of cniCloud, and conducted some experiments to prove the performance and efficiency of this tool. While still at early stage, our experiences have validated cniCloud's potential for querying low-level mobile networks message from a large-scale database. Our project provide a chance for researchers and developers to easily work together on cloud for crowdsourcing low-level cellular network information. In a broader context, cniCloud is designed as a sharing platform for researchers and developers. More community efforts are needed to share the cellular data, and develop advanced query interfaces and algorithms.